%% file: m33.tex

\def\singlespace{\baselineskip 12pt \lineskip 1pt \parskip 2pt plus 1 pt}

\singlespace
\magnification=\magstep1

\raggedbottom


\def\refto#1{$^{#1}$} 
\def\ref#1{ref.~#1} 
\def\Ref#1{#1} 
\gdef\refis#1{\item{#1.\ }} 
\def\beginparmode{\endmode
  \begingroup \def\endmode{\par\endgroup}}
\let\endmode=\par
\def\body{\beginparmode}
\def\head#1{ 
  \goodbreak\vskip 0.5truein 
  {\centerline{\bf{#1}}\par}
   \nobreak\vskip 0.25truein\nobreak}
\def\references 
  {\head{References}            
   \beginparmode
   \frenchspacing \parindent=0pt \leftskip=1truecm
   \parskip=8pt plus 3pt \everypar{\hangindent=\parindent}}
\def\endreferences{\body}

\font\lgh=cmbx10 scaled \magstep2

\def\hb{\hfil\break}

\def\ale{\mathrel{\hbox{\rlap{\hbox{\lower4pt\hbox{$\sim$}}}\hbox{$<$}}}}
\def\age{\mathrel{\hbox{\rlap{\hbox{\lower4pt\hbox{$\sim$}}}\hbox{$>$}}}}

\input reforder.tex

\input citmac.tex

\input psfig.sty


\def\Spectra{1}
\def\Data{2}
\def\Chisq{3}
\def\Msigma{4}


\line{{\bf SCIENCE\  VOL 293\  19 JULY 2001}\hb}
\smallskip
\line{www.sciencemag.org/cgi/content/abstract/1063896\hb}
\bigskip\bigskip
\hrule
\bigskip
\line{\lgh No Supermassive Black Hole in M33?$^{\bf 1}$\hb}
\footnote{}{$^1$Submitted to {\bf Science} 28 June 2001; 
accepted 11 July 2001; published 19 July 2001.
Online version at {\bf Science Express} (www.sciencexpress.org).
Print version appears in the 10 August 2001 issue of {\bf Science}.
Citations to this paper should be in the following format: 
D. Merritt, L. Ferrarese, C. Joseph, {\bf Science}, 19 July 2001 
(10.1126/science.1063896).}

\bigskip

\def\fun#1#2{\lower3.6pt\vbox{\baselineskip0pt\lineskip.9pt
  \ialign{$\mathsurround=0pt#1\hfil##\hfil$\crcr#2\crcr\sim\crcr}}}
\def\lap{\mathrel{\mathpalette\fun <}}
\def\gap{\mathrel{\mathpalette\fun >}}
\def\msun{M_{\odot}}
\def\lsun{L_{\odot}}
\def\mh{M_{\bullet}}
\def\ms{M_{\bullet}-\sigma}

\def\Sec{\hbox{${}^{\prime\prime}$\llap{.}}}
\def\deg{\hbox{${}^\circ$}}

\def\sec{\hbox{${}^{\prime\prime}$}}
\def\kms{km s$^{-1}$}

\line{David Merritt, Laura Ferrarese, Charles L. Joseph\hb}
\smallskip
\line{Department of Physics and Astronomy, Rutgers University, New Brunswick, NJ 08854\hb}
\bigskip
\hrule
\bigskip

\bigskip

\noindent{\bf 
We observed the nucleus of M33, the third-brightest galaxy in the
Local Group, with the Space Telescope Imaging Spectrograph at a
resolution at least a factor of 10 higher than previously obtained.
Rather than the steep rise expected within the radius of gravitational
influence of a supermassive black hole, the random stellar velocities
showed a decrease within a parsec of the center of the galaxy.  The
implied upper limit on the mass of the central black hole is only
$3000 \msun$, about three orders of magnitude lower than the
dynamically-inferred mass of any other supermassive black
hole. Detecting black holes of only a few thousand solar masses is
observationally challenging but is critical for establishing how
supermassive black holes relate  to their host galaxies and which
mechanisms influence the formation and evolution of both.   
}

\bigskip
\bigskip


At a distance of 850 kpc from Earth, M33 is  classified (1) as a
late-type ScII-III spiral, consistent  with its almost nonexistent
bulge (2,3).  The nucleus of M33 is very compact, reaching a stellar
central mass density of several million solar masses per cubic parsec
(4,5), larger than that of any  globular cluster.  While such high
nuclear  densities might be expected in the presence of 
a  supermassive black hole (SMBH) (6), ground-based data
show no evidence for a central rise in stellar 
velocities that would indicate the presence of a compact massive
object in the nucleus (4).

M33 was observed on 12 February 1999 with the Space Telescope Imaging
Spectrograph (STIS) on the Hubble Space Telescope (HST).  Three sets of two
long-slit spectra each, for a total exposure time  of 7380 seconds,
were obtained using the G750M grating centered on the CaII  absorption
triplet near 8561 \AA~ (1 \AA~ corresponds to 10$^{-10}$ meters), covering 19.6
km  s$^{-1}$ pixel$^{-1}$.  The pixel scale is 0\Sec05 with a spatial
resolution of 0\Sec115 at 8561 \AA.   While the two spectra in  each
set were obtained at the same position  to facilitate removal  of
cosmic ray events, the nucleus was moved along the slit by 0\Sec216
between each consecutive set.  This dithering procedure allows for
optimal correction of residual variations in the detector sensitivity
as  well as identification and removal of malfunctioning pixels. The
calibration steps followed the standard procedure (7) adopted for
STIS  observations of the nucleus of M32.

\centerline{\psfig{file=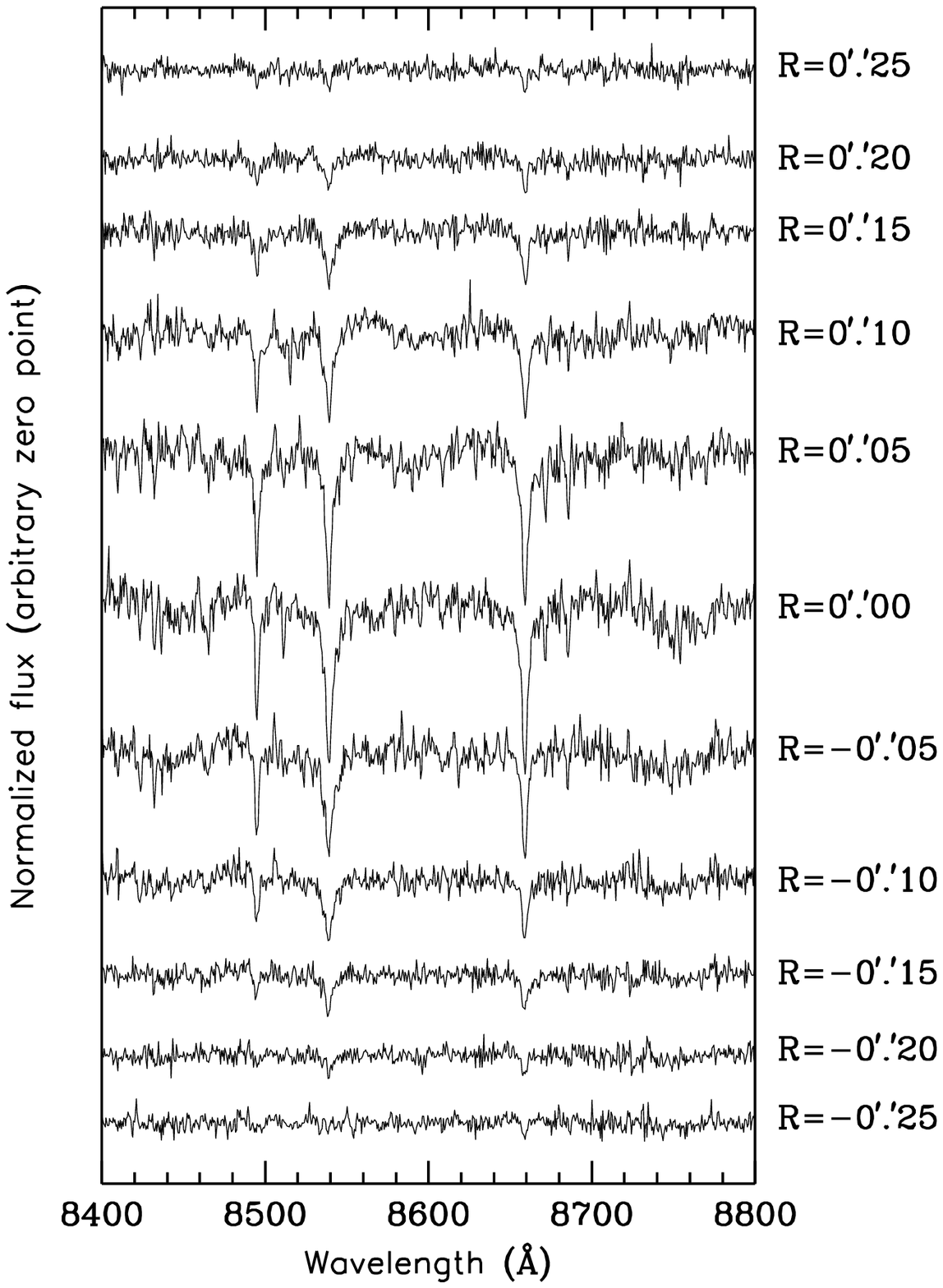,width=6.0in,angle=0}}
\noindent{\bf Figure \Spectra.}

HST/STIS spectra of M33 extracted at the location of the nucleus of
the galaxy (central  row, marked $R=$0\Sec00) and up to 0\Sec25 ($1.0$
pc) off-nucleus.   The spectra have been shifted by an arbitrary
amount in the vertical  direction but the same scale has been
maintained.  The decrease in the  velocity dispersion of the CaII
absorption lines (8561 \AA) at the central positions is visible.

The observed spectrum (Fig. 1) at every resolution element is the
convolution  of the spectra of individual stars with the line-of-sight
velocity distribution (LOSVD) of the stellar ensemble; the latter
contains information about the mean and random velocities of stars in
the nucleus, projected along the line of sight through the galaxy.  As
a typical stellar spectrum we adopted that of the  K0 III giant star
HD7615 which was observed with the same instrumental configuration.
LOSVDs were extracted from the STIS spectra using the Maximum
Penalized Likelihood (MPL) algorithm (8) and represented in terms of
Gauss-Hermite series (9) at every slit position. The mean velocity
$V_0$ and velocity dispersion $\sigma_0$ are given (modulo a standard
correction applied to $V_0$ (9)) by the first two terms in the
Gauss-Hermite expansion.

Neither $V_0$ nor $\sigma_0$ show the rise  that would be expected to
occur whenever a black hole of mass $\mh$ significantly influences the
motion of the stars, i.e.  within a distance from the center of the
galaxy
$$
r_h \equiv {G\mh\over\sigma^2} \approx 0.028''\left({\mh\over
10^4\msun}\right) \left({\sigma\over 20\ {\rm km\ s}^{-1}}\right)^{-2}
\eqno(1) 
$$ 
as seen at the distance of M33. In particular, the central velocity
dispersion is $24\pm 3$ \kms, significantly {\it lower} than the
dispersion of $\sim 35\pm 5$ \kms~ at $\pm 0.3''\approx 1.2$  pc.
Given the $\sim 0.05''$  resolution of STIS, equation (1) implies
$\mh\lap 10^4\msun$ .

The predicted  velocities, however,  depend not  only on $\mh$, but
also on the gravitational potential due to the stars and on the form
of the stellar orbits around the putative black hole (for example are
the stellar orbits eccentric or circular). Therefore, a more rigorous
upper limit to the mass of a possible black hole  can be derived by
constructing realistic dynamical models of the M33 nucleus, which we
assume to be spherical based on its projected circular shape
(5). While a face-on disk would also project circular isophotes,  the
position angle and ellipticity of the nucleus and the outer disk are
considerably different, implying that the former is unlikely to be a
simple extension of the latter. Furthermore, the properties of the M33
nucleus are consistent with those observed for globular clusters (4) -
the prototypical spherical systems. The luminosity density (2,5,10) was
represented by
$$ \nu(r) = \nu_0 \left({r\over a}\right)^{-\gamma}
\left(1+{r\over a}\right)^{\gamma-4}. \eqno(2) 
$$  
In the above equation, $a$ is the distance from the center of the
galaxy at which the luminosity density is a fraction (in our case, one
quarter) of the central value $\nu_0$, while $\gamma$ defines the
gradient in the luminosity density for $r \ll a$. We adopt $\gamma=2$
and $a=1$ kpc (5); $\nu_0$ was fixed by requiring the luminosity
within $1$ pc of the center to be $1.0\times 10^6\lsun$. 
The  gravitational potential due to the stars was derived
from $\nu(r)$ and  from Poisson's equation as a function of the
parameter $M/L$, the ratio  of stellar mass to $V-$ band luminosity
expressed in solar units.   Past studies found $M/L\lap 0.5$
suggestive of a young stellar  population (4,11).

\centerline{\psfig{file=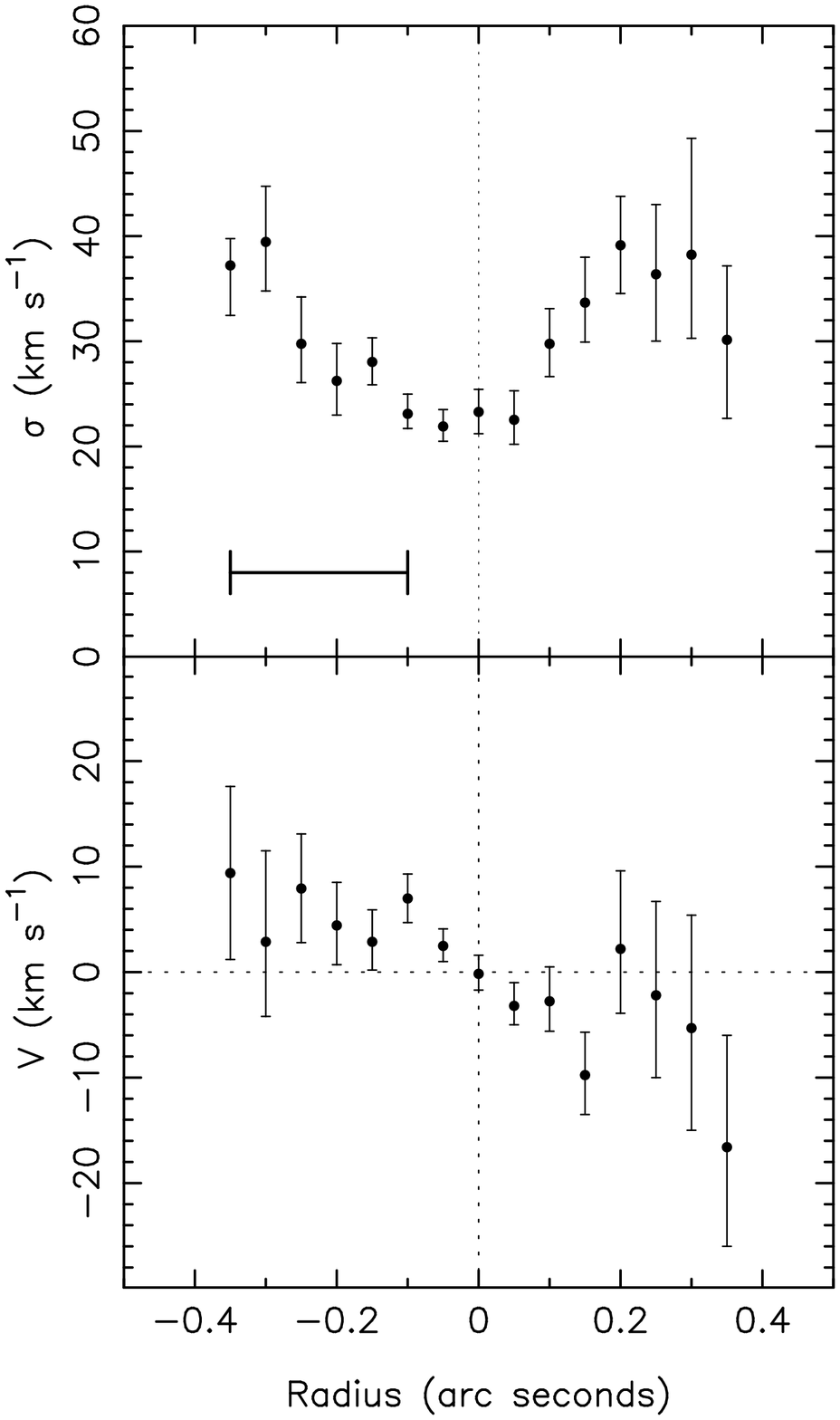,width=5.0in,angle=0}}
\noindent{\bf Figure \Data.}
The stellar rotation curve (lower panel) and velocity dispersion
profile (upper panel) of stars in the nucleus of M33, derived from the
STIS  spectra as described in the text.  One-sigma error bars are
shown. Also shown is the angular scale corresponding to 1 parsec at
the distance of M33.

For a dynamically relaxed stellar system in a spherical gravitational
potential, the number of stars which occupy a given location in
velocity-position space, may be any non-negative function
$f(E,l)$ of the orbital energy $E$ and angular momentum $l$.  We used
standard techniques (12) to find  the $f$ which best reproduced the
kinematical data given $\mh$ and  $M/L$, subject to the constraint
that the integral of $f$ over velocities reproduced the assumed
luminosity density $\nu(r)$.  The velocities
predicted  by $f$ were projected onto the plane of the sky
and convolved with the  HST point spread function (PSF)  and the STIS
slit in order to allow direct comparison with the observed velocities.  
We emphasize that a two-integral ($f(E,L)$) model for a
spherical system, like the model adopted here, permits precisely as
much flexibility in the orbital distribution  as a three-integral
model for an axisymmetric system.  In other words, the constraints on
the mass of the M33 SMBH derived from a three-integral modelling code
would be precisely the same as those found here unless the underlying
stellar potential were assumed to be significantly nonspherical.

In the absence of any additional constraints on $f$, we found that
even black holes with  $\mh\gap 50000\msun$ could be made
consistent with the data.   However the $f$'s corresponding to these
large values of $\mh$ were always  found to be physically
unreasonable: the stellar orbits changed  suddenly from nearly
circular at $r\gap 0.1$ pc to nearly radial at  $r\lap 0.1$ pc,
causing the projected velocity dispersion to drop sharply at a  radius
corresponding to the angular size of the STIS PSF before rising  again
near the black hole.  After convolution with the PSF, the
observed velocities in these solutions therefore remained low even
when $\mh$ was large.  To avoid such unphysical behavior, the
solutions for $f$ were regularized (12), i.e.  forced to  be smooth;
the regularization parameter was chosen to be just large  enough to
suppress unphysical features on the scale of the PSF.

Even after regularization, reasonable fits to the data with  $\mh\gap
10000\msun$ could still be found for values of the stellar mass to
light ratio lower than 0.1 $\msun/L_{\odot}$.  This can be understood
qualitatively as follows. The gravitational potential is
defined by the joint contribution of the stars and the central black
hole. Decreasing the stellar mass-to-light ratio has the effect of
diminishing the stellar contribution to the central potential; to
compensate, the mass of the black hole needs to be increased
proportionally.  Decreasing $M/L$ also has the more subtle effect of
requiring the stellar orbits to become predominantly radial at large
distances. As a result, the predicted line-of-sight  velocity dispersion drops
suddenly just outside of the fitted region, contrary to what is observed:
the velocity dispersion in the M33 bulge appears to remain high,
$\sigma\approx 34$ \kms, within $R\approx$ 80\sec (13).  By
forcing the rms line-of-sight velocity dispersion to be greater than
$30$ \kms\ in the radial range 0\Sec5 $< R <$ 20\sec, such unphysical solutions were excluded.

\centerline{\psfig{file=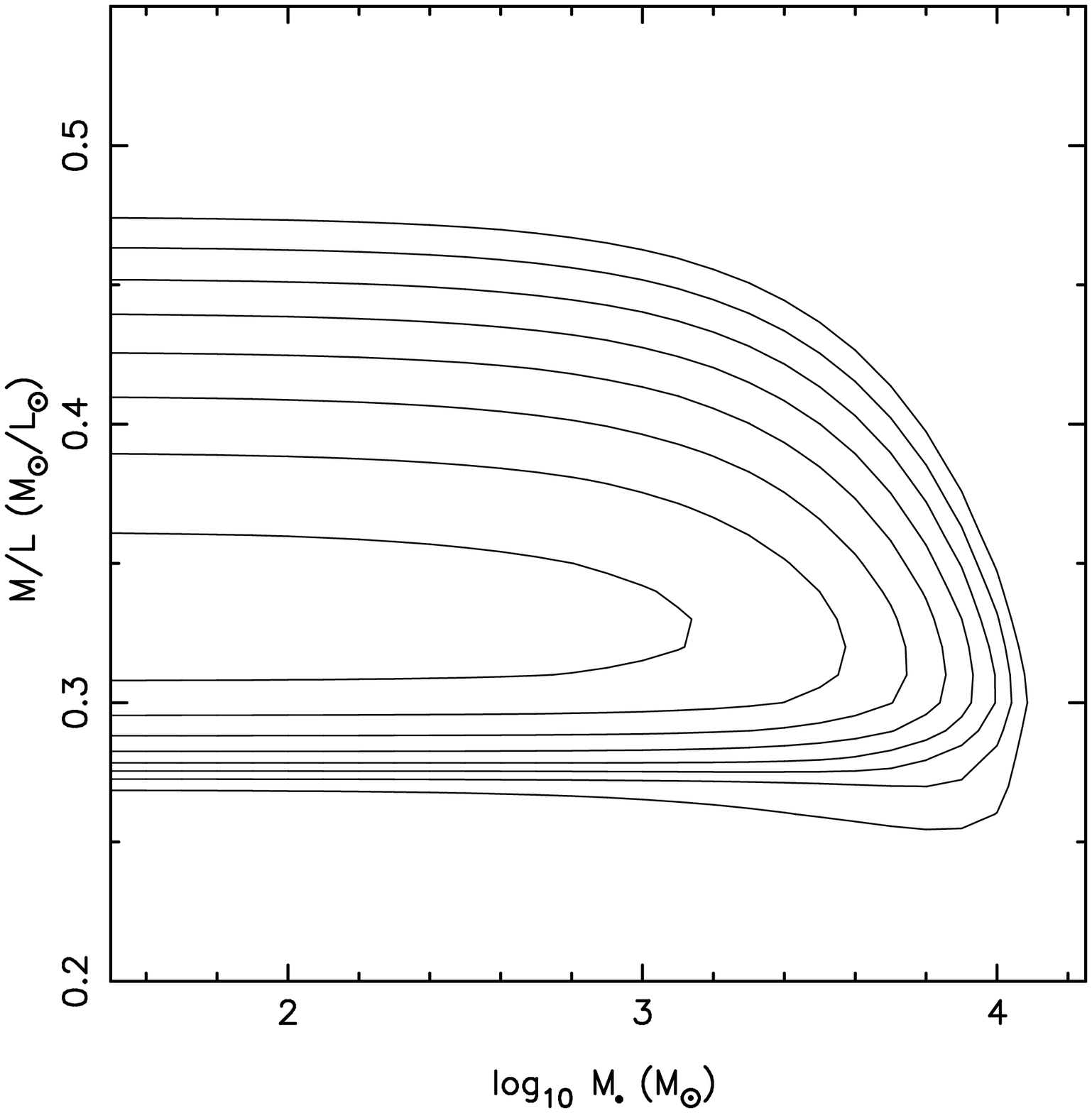,width=5.in,angle=0}}
\noindent{\bf Figure  \Chisq.}
Contours of constant ${\tilde\chi}^2$ measuring the fit of the
dynamical models described in the text to the data of Figure 2.
Horizontal axis is the assumed mass of a central black hole
and vertical axis is the mass-to-light ratio of the stars
in the bulge.
The lowest plotted contour is at ${\tilde\chi}^2=1.4$ and contours 
are separated by $0.2$.

The best-fit $f$ subject to these constraints was computed over a grid
of $(\mh,M/L)$ values (Fig. 3).  For $\mh\lap 1000\msun$ and $M/L\approx
0.35$,  the mass of the putative black hole is too small to
significantly affect  the observable velocities and the quality of the
fit is independent of $\mh$.  As $\mh$ is increased above $\sim
2000\msun$, ${\tilde\chi}^2$  increases as the rise in the central
value of $\sigma$  provides a  progressively worse fit to the data at
$R\lap 0\Sec1$.  Values of $\mh$ as large as $\sim 3000\msun$ imply
${\tilde\chi}^2\approx 1.7$ when only the  data in the innermost two
or three points are compared with the model, and produce a best-fit
model that overpredicts the velocity dispersion at each of the four
innermost points by $\sim$ twice the measurement uncertainty in
$\sigma$. We take $3000\msun$ as our upper limit on $\mh$.   This
value is roughly ten times smaller that the value  $\sim5 \times
10^4\msun$ inferred from ground-based data (4).

There is an empirical relation between the  masses of SMBHs and the
properties of their host bulges called the ``$\ms$  relation,'' which
is expressed as (14)
$$ {\mh\over 10^8\msun} \approx 1.3\left({\sigma_c\over 200\ {\rm km\
s}^{-1}} \right)^{\alpha} \eqno(3)
$$
with $\alpha=4.80\pm 0.54$.  Here $\sigma_c$ is the stellar velocity
dispersion measured within an  aperture of radius $r_e/8$ centered on
the nucleus and $r_e$ is the  projected radius containing $1/2$ of the
light of the bulge.  Because $r_e$  is not well measured for the M33
bulge (estimates range from $0.5$  kpc (2) to $2$ kpc (10)), and the
dependence of  $\sigma$ on radius is not known accurately outside of
the central  $\sim 1$ pc, we conservatively adopt $21\ {\rm km\
s}^{-1}\le\sigma_c\le 34\ {\rm km\ s}^{-1}$, the range of values
measured between $0\sec$ and $80\sec$ (13).  The $\ms$  relation then
predicts a mass in the range  $2600\le\mh/\msun\le 26300$, 
consistent with our upper limit. In contrast, the shallower $\ms$
relation proposed by Gebhardt et  al. (15) would imply  $\mh\gap
2.5\times10^{4} \msun$, which is not consistent with the upper
limit on $\mh$ obtained here. 

The $\ms$ relation is used to  study SMBH demographics and constrain
models of black hole formation  and evolution (16,17,18).  The
low-mass end of the relation is of particular importance because SMBHs
larger than $\sim 10^{6}$ M$_{\odot}$ are believed to originate
through physical processes different from those regulating the
formation of smaller mass black holes (19,20). However, all
supermassive black holes detected so far have masses $\mh >
10^{6}\msun$ (Fig 4). Evidence for ``intermediate-mass black
holes'' (IMBHs),  with masses in the range
$10^2\msun\lap\mh\lap10^5\msun$,  is so far only circumstantial and
relies on speculations  concerning the nature of the super-luminous
off-nuclear X-ray sources  (ULXs) detected in a number of starburst
galaxies (21,22). The connections between the possible black hole in
M33 and ULXs is tantalizing: the  M33 nucleus itself  contains the
brightest ULX in the Local  Group (23), and has optical and
near-infrared colors and spectra consistent with those of a young
cluster (3) with  size and mass similar to those measured for the
cluster containing  the brightest of the M82 ULXs (5,24).  However,
because our upper limit on $\mh$ in M33 is consistent with  the $\ms$
relation as defined by much brighter galaxies,  we cannot yet conclude
that the presence of a black hole in M33 would require a different
formation mechanism from that of the SMBHs detected in other galaxies.

\centerline{\psfig{file=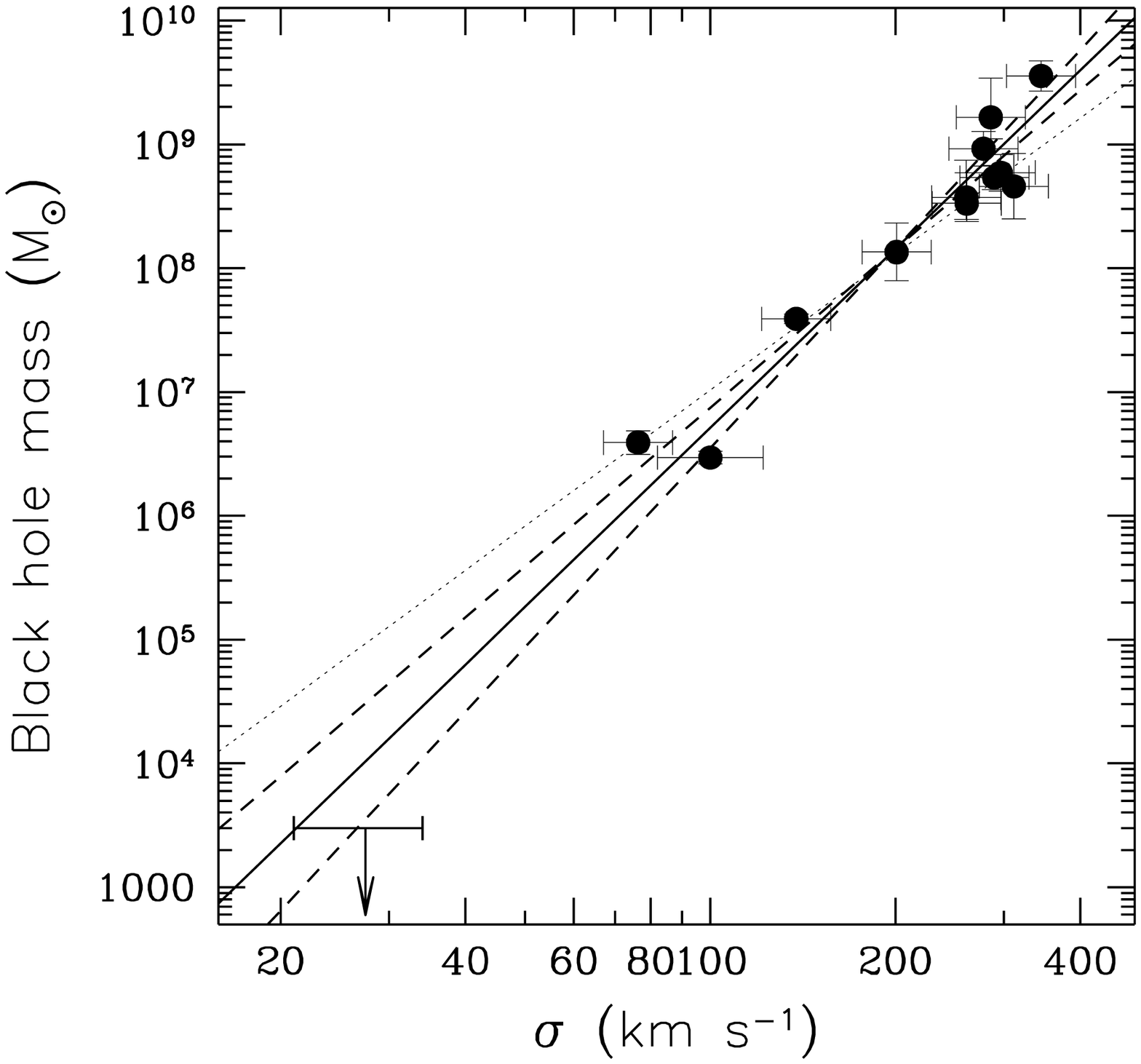,width=5.5in,angle=0}}
\noindent{\bf Figure  \Msigma.}
The thick solid line represents the $\ms$~ relation as derived by
Ferrarese \&  Merritt (14), with 1$-\sigma$ confidence limits on the
slope shown by the dashed lines. The upper limit for  the black hole
mass in M33 (shown by the arrow) is consistent with this relation but
inconsistent with the shallower relation advocated by Gebhardt et
al. (15) and shown by the thin dotted line.

\vfill\eject

\centerline {\bf References}

\bigskip

\refis{vandenbergh99} van den Bergh, S.  The local group of galaxies.  {\it 
Ann.  Rev.  astron.  Ap.} {\bf 9}, 273--318 (1999).

\refis{Minniti93} Minniti, D., Olszewski, E.  W.  \& Rieke, M.  The Bulge 
of M33.  {\it Astrophys.  J.} {\bf 410}, L79--L82 (1993).

\refis{vandenbergh91} van den Bergh, S.  The stellar populations of M33.  
{\it Publ.  Astron.  Soc.  Pacif.} {\bf 103}, 609--622 (1991).

\refis{Kormendy93} Kormendy, J., \&\ McClure, R.  D.  The nucleus of M33 
{\it Astron.  J.} {\bf 105}, 1793--1812 (1993).

\refis{lau98} Lauer, T.  R., Faber, S.  M., Ajhar, E.  A., Grillmair, C.  
J.  \& Scowen, P.  A., M32$\pm 1$.  {\it Astron.  J.} {\bf 116}, 
2263--2286 (1998).  

\refis{vandermarel99} van der Marel, R.  The black hole mass distribution 
in early-type galaxies: cusps in Hubble Space Telescope photometry 
interpreted through adiabatic Black Hole growth.  {\it Astron.  J.} {\bf 
117}, 744--763 (1999).

\refis{Joseph01} Joseph, C., et al.  The nuclear dynamics of M32.  I.  Data 
and stellar kinematics.  {\it Astrophys.  J.} {\bf 550}, 668--690 (2001)

\refis{Merritt97} Merritt, D.  Recovering velocity distributions via 
penalized likelihood.  {\it Astron.  J.} {\bf 114}, 228--237 (1997).

\refis{Marel93} van der Marel, R.  P., \& Franx, M.  A new 
method for the identification of non-Gaussian line profiles in elliptical 
galaxies.  {\it Astrophys.  J.} {\bf 407}, 525--539 (1993).

\refis{Regan94} Regan, M.  W., \& Vogel, S.  N.  The near-infrared 
structure of M33.  {\it Astron.  J.} {\bf 434}, 536--545 (1994).

\refis{Takamiya00} Takamiya, T., Sofue, Y. Radial distribution of the 
mass-to-luminosity ratio in spiral galaxies and massive dark cores.
{\it Astrophys. J.} {\bf 534}, 670--683 (2000)

\refis{Merritt93} Merritt, D.  Dynamical mapping of hot stellar systems.  
{\it Astrophys.  J.} {\bf 413}, 79-94 (1993).

\refis{Minniti96} Minniti, D.  Velocities of supergiants in the bulge of M 
33.  {\it Astron.  Astrophys.} {\bf 306}, 715--720 (1996).

\refis{Ferrarese00} Ferrarese, L., \& Merritt, D.  A fundamental relation 
between supermassive black holes and their host galaxies.  {\it Astrophys.  
J.} {\bf 539}, L9--L12 (2000).

\refis{Gebhardt00} Gebhardt, K.  Bender, R., Bower, G., Dressler, A., 
Faber, S.  M., Filippenko, A.  V., Green, R., Grillmair, C., Ho, L.  C..  
Kormendy, J., Lauer, T.  R., Magorrian, J., Pinkney, J., Richstone, D., 
Tremaine, S.  A relationship between nuclear black hole mass and galaxy 
velocity dispersion.  {\it Astrophys.  J.} {\bf 539}, L13--L16 (2000).



\refis{Haehnelt00} Haehnelt, M.  G., \&\ Kauffmann, G.  The correlation 
between black hole mass and bulge velocity dispersion in hierarchical 
galaxy formation models.  {\it Mon.  Not.  R.  astron.  Soc.} {\bf 318}, 
L35--L38 (2000).

\refis{Ciotti01} Ciotti, L., \&\ van Albada, T.  S.  The $\ms$~ relation as 
a constraint on the formation of elliptical galaxies.  Accepted by {\it 
Astrophys.  J.} (astro-ph/0103336) (2001)

\refis{Merritt01b} Merritt, D., \& Ferrarese, L.  Black Hole demographics 
from the $\ms$ relation.  {\it Mon.  Not.  R.  astron.  Soc.} {\bf 320}, 
L30--L34 (2001).

\refis{Haehnelt98} Haehnelt, M.  G., Natarajan, P., \&\ Rees, M.  J.  The 
distribution of supermassive black holes in the nuclei of nearby galaxies.  
{\it Mon.  Not.  R.  astron.  Soc.} {\bf 308}, 77--81 (1998).

\refis{Miller01} Miller, M.  C., \&\ Hamilton, D.  P.  Production of 
intermediate-mass black holes in globular clusters.  Submitted to {\it Mon.  
Not.  R.  astron.  Soc.} (astro-ph/0106188) (2001).

\refis{Matsumoto01}Matsumoto, H., Tsuru, T. G., Koyama, K., Awaki, H., 
Canizares, C. R.; Kawai, N., Matsushita, S., Kawabe, R. 
Discovery of a luminous, variable, off-center source in the nucleus of 
M82 with the Chandra High-Resolution Camera.
{\it Astrophys. J.} {\bf 547}, L25--L28 (2001)

\refis{Fabbiano01} Fabbiano, G., Zezas, A., Murray, S.S. Chandra 
observations of ``the Antennae'' galaxies (NGC 4038/39).  Accepted by
{\it Astrophys. J.} (astro-ph/0102256) (2001)

\refis{Long81} Long, K.S., Dodorico, S., Charles, P.A., \&\ Dopita, 
M.A. Observations of the X-ray sources in the nearby SC galaxy M33.
{\it Astrophys. J.} {\bf 246}, L61--L64 (1981)

\refis{Ebisuzaki01} Ebisuzaki, T., Makino, J., Tsuru, T. G., Funato, 
Y., Portegies Zwart, S., Hut, P., McMillan, S., Matsushita, S., 
Matsumoto, H., \&\ Kawabe, R. Missing link found? --- the ``runaway'' path 
to supermassive black holes. Submitted to {\it Astrophys. J.} 
(astro-ph/0106252) (2001).

\refis{Acknowledgment} This work was supported by the National Science 
Foundation through grant 4-21911, and by the National Aeronautics and
Space Administration through grants 4-21904 and NAG5-8693.

\endreferences
\vfill\eject
\endmode

\bye

%% file: reforder.tex
\catcode`@=11
\newcount\r@fcount \r@fcount=0
\newcount\r@fcurr
\immediate\newwrite\reffile
\newif\ifr@ffile\r@ffilefalse
\def\w@rnwrite#1{\ifr@ffile\immediate\write\reffile{#1}\fi\message{#1}}

\def\writer@f#1>>{}
\def\referencefile{
  \r@ffiletrue\immediate\openout\reffile=\jobname.ref%
  \def\writer@f##1>>{\ifr@ffile\immediate\write\reffile%
    {\noexpand\refis{##1} = \csname r@fnum##1\endcsname = %
     \expandafter\expandafter\expandafter\strip@t\expandafter%
     \meaning\csname r@ftext\csname r@fnum##1\endcsname\endcsname}\fi}%
  \def\strip@t##1>>{}}

\def\citeall#1{\xdef#1##1{#1{\noexpand\cite{##1}}}}
\def\cite#1{\each@rg\citer@nge{#1}}	

\def\each@rg#1#2{{\let\thecsname=#1\expandafter\first@rg#2,\end,}}
\def\first@rg#1,{\thecsname{#1}\apply@rg}	
\def\apply@rg#1,{\ifx\end#1\let\next=\relax
\else,\thecsname{#1}\let\next=\apply@rg\fi\next}

\def\citer@nge#1{\citedor@nge#1-\end-}	
\def\citer@ngeat#1\end-{#1}
\def\citedor@nge#1-#2-{\ifx\end#2\r@featspace#1 
  \else\citel@@p{#1}{#2}\citer@ngeat\fi}	
\def\citel@@p#1#2{\ifnum#1>#2{\errmessage{Reference range #1-#2\space is bad.}%
    \errhelp{If you cite a series of references by the notation M-N, then M and
    N must be integers, and N must be greater than or equal to M.}}\else%
 {\count0=#1\count1=#2\advance\count1 by1\relax\expandafter\r@fcite\the\count0,%
  \loop\advance\count0 by1\relax
    \ifnum\count0<\count1,\expandafter\r@fcite\the\count0,%
  \repeat}\fi}

\def\r@featspace#1#2 {\r@fcite#1#2,}	
\def\r@fcite#1,{\ifuncit@d{#1}
    \newr@f{#1}%
    \expandafter\gdef\csname r@ftext\number\r@fcount\endcsname%
                     {\message{Reference #1 to be supplied.}%
                      \writer@f#1>>#1 to be supplied.\par}%
 \fi%
 \csname r@fnum#1\endcsname}
\def\ifuncit@d#1{\expandafter\ifx\csname r@fnum#1\endcsname\relax}%
\def\newr@f#1{\global\advance\r@fcount by1%
    \expandafter\xdef\csname r@fnum#1\endcsname{\number\r@fcount}}

\let\r@fis=\refis			
\def\refis#1#2#3\par{\ifuncit@d{#1}
   \newr@f{#1}%
   \w@rnwrite{Reference #1=\number\r@fcount\space is not cited up to now.}\fi%
  \expandafter\gdef\csname r@ftext\csname r@fnum#1\endcsname\endcsname%
  {\writer@f#1>>#2#3\par}}

\def\ignoreuncited{
   \def\refis##1##2##3\par{\ifuncit@d{##1}%
     \else\expandafter\gdef\csname r@ftext\csname r@fnum##1\endcsname\endcsname%
     {\writer@f##1>>##2##3\par}\fi}}

\def\r@ferr{\endreferences\errmessage{I was expecting to see
\noexpand\endreferences before now;  I have inserted it here.}}
\let\r@ferences=\references
\def\references{\r@ferences\def\endmode{\r@ferr\par\endgroup}}

\let\endr@ferences=\endreferences
\def\endreferences{\r@fcurr=0
  {\loop\ifnum\r@fcurr<\r@fcount
    \advance\r@fcurr by 1\relax\expandafter\r@fis\expandafter{\number\r@fcurr}%
    \csname r@ftext\number\r@fcurr\endcsname%
  \repeat}\gdef\r@ferr{}\endr@ferences}


\let\r@fend=\endpaper\gdef\endpaper{\ifr@ffile
\immediate\write16{Cross References written on []\jobname.REF.}\fi\r@fend}

\catcode`@=12

\citeall\refto		
\citeall\ref		%
\citeall\Ref		%

%% file: citmac.tex
\def\singlespace{\baselineskip 12pt \lineskip 1pt \parskip 2pt plus 1 pt}

\def\today{\number\day\enspace
     \ifcase\month\or January\or Febuary\or March\or April\or May\or
     June\or July\or August\or September\or October\or
     November\or December\fi \enspace\number\year}
\def\clock{\count0=\time \divide\count0 by 60
    \count1=\count0 \multiply\count1 by -60 \advance\count1 by \time
    \number\count0:\ifnum\count1<10{0\number\count1}\else\number\count1\fi}
\footline={\hss -- \folio\ -- \hss}

\def\deg{\ifmmode^\circ\else$^\circ$\fi}
\def\solar{\ifmmode_{\mathord\odot}\else$_{\mathord\odot}$\fi}
\def\jref#1 #2 #3 #4 {{\par\noindent \hangindent=3em \hangafter=1 
      \advance \rightskip by 5em #1, {\it#2}, {\bf#3}, #4.\par}}
\def\ref#1{{\par\noindent \hangindent=3em \hangafter=1 
      \advance \rightskip by 5em #1.\par}}
\newcount\eqnum
\def\nexteq{\global\advance\eqnum by1 \eqno(\number\eqnum)}
\def\lasteq#1{\if)#1[\number\eqnum]\else(\number\eqnum)\fi#1}
\def\preveq#1#2{{\advance\eqnum by-#1
    \if)#2[\number\eqnum]\else(\number\eqnum)\fi}#2}

\def\tableheight{\vrule width 0pt height 8.5pt depth 3.5pt}
{\catcode`|=\active \catcode`&=\active 
    \gdef\tabledelim{\catcode`|=\active \let|=\vbar
                     \catcode`&=\active \let&=\nobar} }
\def\table{\begingroup
    \def\twidth{\hsize}
    \def\tablewidth##1{\def\twidth{##1}}
    \def\defaultheight{\vrule width 0pt height 8.5pt depth 3.5pt}
    \def\heightdepth##1{\dimen0=##1
        \ifdim\dimen0>5pt 
            \divide\dimen0 by 2 \advance\dimen0 by 2.5pt
            \dimen1=\dimen0 \advance\dimen1 by -5pt
            \vrule width 0pt height \the\dimen0  depth \the\dimen1
        \else  \divide\dimen0 by 2
            \vrule width 0pt height \the\dimen0  depth \the\dimen0 \fi}
    \def\spacing##1{\def\defaultheight{\heightdepth{##1}}}
    \def\nextheight##1{\noalign{\gdef\tableheight{\heightdepth{##1}}}}
    \def\end{\cr\noalign{\gdef\tableheight{\defaultheight}}}
    \def\zerowidth##1{\omit\hidewidth ##1 \hidewidth}    
    \def\hline{\noalign{\hrule}}
    \def\skip##1{\noalign{\vskip##1}}
    \def\bskip##1{\noalign{\hbox to \twidth{\vrule height##1 depth 0pt \hfil
        \vrule height##1 depth 0pt}}}
    \def\header##1{\noalign{\hbox to \twidth{\hfil ##1 \unskip\hfil}}}
    \def\bheader##1{\noalign{\hbox to \twidth{\vrule\hfil ##1 
        \unskip\hfil\vrule}}}
    \def\spanloop{\span\omit \advance\mscount by -1}
    \def\extend##1##2{\omit
        \mscount=##1 \multiply\mscount by 2 \advance\mscount by -1
        \loop\ifnum\mscount>1 \spanloop\repeat \ \hfil ##2 \unskip\hfil}
    \def\vbar{&\vrule&}
    \def\nobar{&&}
    \def\hdash##1{ \noalign{ \relax \gdef\tableheight{\heightdepth{0pt}}
        \toks0={} \count0=1 \count1=0 \putout##1\end 
        \toks0=\expandafter{\the\toks0 &\end} \xdef\piggy{\the\toks0} }
        \piggy}
    \let\e=\expandafter
    \def\putspace{\ifnum\count0>1 \advance\count0 by -1
        \toks0=\e\e\e{\the\e\toks0\e&\e\multispan\e{\the\count0}\hfill} 
        \fi \count0=0 }
    \def\putrule{\ifnum\count1>0 \advance\count1 by 1
        \toks0=\e\e\e{\the\e\toks0\e&\e\multispan\e{\the\count1}\leaders\hrule\hfill}
        \fi \count1=0 }
    \def\putout##1{\ifx##1\end \putspace \putrule \let\next=\relax 
        \else \let\next=\putout
            \ifx##1- \advance\count1 by 2 \putspace
            \else    \advance\count0 by 2 \putrule \fi \fi \next}   }
\def\tablespec#1{
    \def\vdimens{\noexpand\tableheight}
    \def\tabby{\tabskip=0pt plus100pt minus100pt}
    \def\r{&################\tabby&\hfil################\unskip}
    \def\c{&################\tabby&\hfil################\unskip\hfil}
    \def\l{&################\tabby&################\unskip\hfil}
    \edef\templ{\noexpand\vdimens ########\unskip  #1 
         \unskip&########\tabskip=0pt&########\cr}
    \tabledelim
    \edef\body##1{ \vbox{
        \tabskip=0pt \offinterlineskip
        \halign to \twidth {\templ ##1}}} }

\newbox\grsign \setbox\grsign=\hbox{$>$}
\newdimen\grdimen \grdimen=\ht\grsign
\newbox\laxbox \newbox\gaxbox
\setbox\gaxbox=\hbox{\raise.5ex\hbox{$>$}\llap
	{\lower.5ex\hbox{$\sim$}}}\ht1=\grdimen\dp1=0pt
\setbox\laxbox=\hbox{\raise.5ex\hbox{$<$}\llap
	{\lower.5ex\hbox{$\sim$}}}\ht2=\grdimen\dp2=0pt

\def\uJy{\ifmmode{\,\mu{\rm Jy}}\else$\,{\mu{\rm Jy}}$\fi}
\def\mJy{\ifmmode{\,{\rm mJy}}\else${\,{\rm mJy}}$\fi}
\def\MHz{\ifmmode{\,{\rm MHz}}\else{$\,{\rm MHz}$}\fi}
\def\GHz{\ifmmode{\,{\rm GHz}}\else{$\,{\rm GHz}$}\fi}
\def\solar{\ifmmode_{\mathord\odot}\else$_{\mathord\odot}$\fi}
\def\Msolar{\ifmmode{\, {\rm M\solar}}\else{${\, {\rm M\solar}}$}\fi}
\def\Rsolar{\ifmmode{\, {\rm R\solar}}\else{${\, {\rm R\solar}}$}\fi}
\def\kms{\ifmmode{\,{\rm km\,s^{-1}}}\else${\,{\rm km\,s^{-1}}}$\fi}
\def\kpc{\ifmmode{\,{\rm kpc}}\else${\,{\rm kpc}}$\fi}
\def\us{\ifmmode{\,\mu{\rm s}}\else$\,{\mu{\rm s}}$\fi}
\def\ms{\ifmmode{\,{\rm ms}}\else$\,{{\rm ms}}$\fi}
\def\y{\ifmmode{\,{\rm y}}\else$\,{\rm y}$\fi}
\def\h{\ifmmode{^{\rm h}}\else$^{\rm h}$\fi}
\def\m{\ifmmode{^{\rm m}}\else$^{\rm m}$\fi}
\def\s{\ifmmode{^{\rm s}}\else$^{\rm s}$\fi}
\def\Lmin{\ifmmode{L_{min}}\else{$L_{min}$}\fi}